# Organic film thickness influence on the bias stress instability in Sexithiophene Field Effect Transistors

F.V. Di Girolamo[1], C. Aruta[1], M. Barra[1], P. D'Angelo[1,2] and A. Cassinese[1]

[1]CNR-INFM COHERENTIA and Department of Physics Science, University of Naples Federico II, Piazzale Tecchio, 80125 Naples, Italy

[2] CNR-ISMN via Pietro Gobetti, 40129 Bologna, Italy

**Abstract**: In this paper, the dynamics of bias stress phenomenon in Sexithiophene (T6) Field Effect Transistors (FETs) has been investigated. T6 FETs have been fabricated by vacuum depositing films with thickness from 10 nm to 130 nm on Si/SiO$_2$ substrates. After the T6 film structural analysis by X-Ray diffraction and the FET electrical investigation focused on carrier mobility evaluation, bias stress instability parameters have been estimated and discussed in the context of existing models. By increasing the film thickness, a clear correlation between the stress parameters and the structural properties of the organic layer has been highlighted. Conversely, the mobility values result almost thickness independent.

Corresponding author: Flavia Viola Di Girolamo

Dipartimento Scienze Fisiche, Università

di Napoli Federico II and CNR-INFM Coherentia,

Piazzale Tecchio, 80, 80125, Naples (Italy)

Ph: +39-(0)817682548, Fax: +39-(0)812391821

E-mail: fdigirolamo@na.infn.it

# 1. INTRODUCTION

In last two decades, organic semiconductors have received increasing attention for their peculiar features, such as low cost processing techniques and potential use of large area and flexible substrates. Consequently, a wide variety of electronic devices, like memories[1], RFIDs[2], sensors[3,4], based on this new class of materials, has been experimentally demonstrated. In particular, Organic Thin Film Transistors (OTFTs) have been the subject of intense interest and investigation, showing to exceed the charge carrier mobility performances of hydrogenated amorphous silicon $\alpha$-Si:H TFTs[5]. Anyway, OTFTs are generally characterized by current instability, whose main manifestations are hysteresis in current-voltage curves, threshold voltage shift and current decay with time under fixed operation conditions[6--10]. Indeed, these effects are basically related to continuous Gate voltage application (bias stress), while the corresponding influence of Drain-Source voltage seems to be negligible[9]. Provided that, it is obvious that current instability effects can play an important role in defining the overall OTFTs electrical performances and in making possible their practical application in complex electronic circuitry. Hence, at this development stage, the understanding of physical mechanisms underlying the bias stress phenomena appears as a crucial step towards the actual achievement of this new technology. So far, many different experimental approaches have been considered in the attempt to gain deeper insights on the current instability issues. In detail, the related influence of different factors, such as dielectric surface treatments[11] or environmental conditions with the presence of water and oxygen[12--14], has been assessed. In similar way, light was also shown to affect the current time evolution, accelerating the threshold voltage recovery after bias stress occurrence[15] and magnifying the hysteresis evidence in transfer curves[16]. Commonly, all these current instability mechanisms are explained by the presence of trapping sites, with energies close to the conduction or valence levels and the capability to immobilize the charge carriers. Anyway, their origin and physical location inside the semiconductor structure are still under debate. Recent reports suggest that the traps involved in bias stress effects are basically related to the hydrophobic properties of the dielectric/semiconductor interface, with the subsequent absorption of water molecules acting as very effective trapping centres[9]. According to this scenario, the semiconductor type and its properties should play a minor role on instability mechanisms. Differently, the experimental findings of other papers, mainly focused on the use of various dielectric barriers[15] and on the effect of active layer coverage[17], do not exclude the influence of the film features on bias stress effects. With the attempt to further address this last issue, in this study appropriate procedures for the determination of stress electrical parameters have been performed to ascertain the influence of the organic layer thickness on Sexithiophene (T6) transistor instability. T6 FET device properties have been analysed by both structural and electrical techniques. Bias stress effects have been assessed by different experimental approaches and particular care has been taken in fitting the experimental data with theoretical models.

## 2. EXPERIMENTAL

Bottom-contact bottom-gate TFTs were fabricated by evaporating T6 films, with thickness ranging between 10 nm and 130 nm, on substrates consisting of a heavily n-doped silicon (Gate contact), a 200 nm thick $SiO_2$ thermally grown layer (dielectric barrier) and interdigitated gold electrodes (Source-Drain contacts). The layout of the test pattern is sketched in Fig.1. Each substrate provides four devices separated in two couples sharing the same geometrical features. For two devices the channel length L is 40 µm, while it is reduced to 20 µm for the other two. Anyway, the corresponding channel widths W are scaled in order to keep fixed the ratio W/L=550. T6 films were evaporated by Knudsen cells in a high vacuum system (base pressure between $10^{-8}$ and $10^{-7}$ mbar); the entire vacuum chamber was heated at the same temperature (90 °C) for 24 hours, in order to assure uniform warming of the substrate. More details about the morphology and the film deposition have been reported elsewhere[18,19]. Before the deposition, test patterns were cleaned by sonication in acetone and ethanol baths, followed by a drying in pure $N_2$ gas. In this work, for all the thickness considered, the growth rate was fixed to about 0.2 Å/sec. For all the depositions, the temperature of the Knudsen cell was kept fixed and all the samples were grown without refilling the cells. This procedure provides a good reproducibility of the film deposition parameters including the surface energy but prevents the growth of thicker films.

## 3. RESULTS AND DISCUSSION

### 3.1 X RAYS CHARACTERIZATION

T6 films have been structurally characterised trough X-ray diffraction measurements. X-ray diffraction (XRD) θ-2θ scans and rocking curves were performed in symmetrical reflection mode at the Cu $K\alpha$ wavelength radiation. The results for different thicknesses between 20 and 130 nm are reported in fig.2.

In the θ-2θ spectrum of the thicker T6 films, only the ($h$00) reflections can be clearly observed. By decreasing the film thickness, these reflections become weaker and weaker. The presence of only the ($h$00) diffraction peaks is a clear indication that films are well structured and aligned with the long crystallographic axis along the direction perpendicular to the substrate surface. The high crystal quality of the T6 films is confirmed in the case of d=130 nm film by the narrow rocking curve of the (12,0,0) reflection, with a Full Width at Half Maximum (FWHM) of about 0.38°. In agreement with the θ-2θ spectra, the rocking curves are wider and wider for the thinner films, thus resulting FWHM=0.82° in the case of the 23 nm thick film. Moreover, it should be noted that for T6 film thicknesses greater than 60 nm the rocking curves at the (20,0,0) reflection are still measurable, showing a narrow FWHM (about 0.5°). The depression of diffraction peaks intensity for lower thicknesses and the corresponding increase of width in rocking curves, both indicate that by increasing the film thickness the crystalline domains tend to widen as a consequence of nucleation processes occurred at low coverage.

### 3.2 CHARGE TRANSPORT ANALYSIS

Electrical measurements were carried out always in darkness, both in vacuum (about $10^{-4}$ mbar) and in air, by using a cryogenic probe station connected to a Keithley 487 picoammeter and a Keithley 2400 voltmeter. The charge transport properties of the T6 FETs have been assessed by measuring the output ($I_{DS}$ vs $V_{DS}$ at different $V_{GS}$) and transfer- ($I_{DS}$ vs $V_{GS}$ at different $V_{DS}$) curves. This analysis was mainly devoted to the determination of the charge carrier mobility µ value. The transfer-curves have been modelled by using the standard MOSFET equations in linear and saturation regimes, respectively:

$$I_{DS} = \frac{W\mu C_{OX}}{L}(V_{GS} - V_{th})V_{DS} \qquad |(V_{GS} - V_{th})| \gg V_{DS} \qquad 1)$$

$$I_{DS} = \frac{W\mu C_{OX}}{2L}(V_{GS} - V_{th})^2 \qquad |(V_{GS} - V_{th})| \leq V_{DS} \qquad 2),$$

where $C_{OX}$ is the dielectric barrier capacitance per unit area and $V_{th}$ is the threshold voltage. Based on these expressions, the field effect mobility can be easily evaluated both in linear and saturation regions by the relations, respectively:

$$\mu_{lin} = \frac{\partial I_{DS}}{\partial V_{GS}} * \frac{L}{WC_{OX}V_{DS}}, \qquad 3)$$

$$\mu_{sat} = \frac{\partial \sqrt{I_{DS}}}{\partial V_{GS}} * \frac{2L}{WC_{OX}} \qquad 4).$$

Two experimental transfer-curves in saturation regime, measured in vacuum and air for the same T6 transistor, are reported in fig.3. This comparison well clarifies the general air effect in reducing the overall FET current and the subsequent carrier mobility. Indeed, depending on the sample, the mobility reduction factor due to the air exposure was experimentally tested to range between 2 and 5. This occurrence can be explained by invoking namely the effect of ambient and moisture doping which increase the trapping centre density and, together with the mobility lowering, tend contemporarily to emphasize the hysteresis in the experimental transfer-curves and to shift $V_{th}$ toward more positive voltages[12]. Hereafter, all reported results are referred to measurements performed in vacuum, where the extrinsic doping effects are minimized. For any investigated device, mobility values have been evaluated by equation 3) and 4). The mobility dependence on the film thickness (between 10 nm and 130 nm) is presented in the inset of fig.3.

As shown, except for d=10 nm, mobility exhibits almost constant values for all thicknesses (ranging from $1.1*10^{-3}$ to $1.6*10^{-2}$ cm$^2$/volt*sec in linear region and from $1.3*10^{-3}$ to $1.8*10^{-2}$ cm$^2$/volt*sec in saturation region). In agreement with previous reports[20], this evidence confirms that the charge carrier transport in these devices is basically an interfacial phenomenon involving only a few nanometer thick region. The small discrepancy between the mobility values extracted in linear ($V_{DS}$=-5 Volt) and saturation ($V_{DS}$=-50 Volt) regimes, together with the linear behaviour of the output curves (not shown here) in the small $V_{DS}$ region, are clear manifestations of a negligible influence of the contact resistances at Drain-Source electrodes. In order to gain more insights into the basic charge transport mechanism of our films, variable temperature transfer-curve measurements have been performed and the mobility temperature dependence has been extracted. Fig.4 reports a typical set of transfer-curves measured at different temperatures from room temperature to 70 K. The inset of fig.4 reveals that mobility is thermally activated and obeys the so-called Arrhenius law:

$$\mu(T) = \mu_0 * e^{-\frac{E_a}{KT}}$$

where $E_a$ is the activation energy. The same experimental occurrence has been verified for all the analyzed samples and activation energy values ranging between 70 and 90 meV have been found[21], without a clear correlation with the film thickness. This type of behaviour agrees very well with the theoretical predictions of the Multiple Trap and Release (MTR) model[22], suggesting that charge transport is a dynamic process ruled by the competition between trapping phenomena, involving energy states localized in the band tail above the valence edge, and subsequent thermal release. According to this model, mobility is basically given by the ratio between free and trapped charge carrier densities, and, under the simplified assumption of a single discrete trap state level at energy $E_T$, Ea represents the energy difference between $E_T$ and the valence band edge.

## 3.3 BIAS STRESS INSTABILITY MEASUREMENTS

Up to date, bias stress phenomena in polycrystalline or amorphous semiconductors have been mainly analyzed by the experimental observations of the threshold voltage shift $V_{th}$ in FET transfer-curves, upon the prolonged application of a Gate voltage. This effect has been widely investigated in α-Si:H TFTs, where it has been studied as a dispersive phenomenon[23--25] and the stretched exponential function has been utilized to describe the threshold voltage time evolution[26,27] :

$$\Delta V_{th}(t) = V_0 \left\{ 1 - \exp\left[-\left(\frac{t}{\tau}\right)^{\beta}\right] \right\} \qquad (5).$$

It is important to outline that in this equation $\Delta V_{th}=V_{th}(t)-V_{th}(0)$ and $V_0=V_{GS}-V_{th}(0)$, being $V_{GS}$ and $V_{th}(0)$ the applied Gate voltage and the threshold voltage at the initial stage, respectively. Consequently, this model foresees that at the infinite time, where $\Delta V_{th}=V_0$, $V_{th}(t)$ has to be equal to the applied $V_{GS}$. More in detail, the stretched exponential law is deduced under the basic assumption that the threshold voltage shift rate $dV_{th}/dt$ is proportional to the variation of the trapped charges $dN_t/dt$, according to the relation:

$$\frac{dV_{th}}{dt} \propto \frac{dN_t}{dt} \propto N_f(t) * \frac{t^{\beta-1}}{\tau^{\beta}} \qquad (6).$$

In (6), $N_f(t)$ is the free-carrier density, $\tau$ is a relaxation time characteristic of the trapping mechanism and the dispersion parameter β is related to the width of the exponential energy distribution of the trapping sites above the valence band edge or below the conduction band edge for p-type and n type semiconductors, respectively. In recent years, independently form the physical origin of the basic mechanisms involved in trapping processes, the threshold voltage shift law expressed by (5) has been demonstrated to be suitable also to give a phenomenological description of stress effects in organic field effect devices[9,16,28] . To this regard, fig.5a reports the results of a typical bias stress experiment performed in our study. The figure represents the time evolution (up to $2*10^4$ sec) of a T6 transistor transfer-curves in linear regime ($V_{DS}$=-5 Volt), under the application of Gate voltage $V_{GS}$=-50 Volt. As shown, the transfer-curves slope does not change with time and consequently the field effect mobility results to be not affected by the bias stress occurrence. This is a general result which perfectly agrees with most experimental evidences reported in literature[7,9]. Conversely, the threshold voltage shift, which can be obtained from the plot of the transfer-curves as the Gate voltage value for which the linear extrapolation to any transfer-curve crosses the $V_{GS}$ axis, is very pronounced.

The $V_{th}$ shift evolution, deduced by the abovementioned criterion, is presented as a function of time in the semi-log plot of fig.5b. After $2*10^4$ sec, $\Delta V_{th}$ is about 16.5V, so by using the equation:

$$N_t = \frac{\Delta V_{th} * C_{ox}}{e}$$

the corresponding density of trapped charges can be estimated to be about $1.8*10^{12}$ cm$^{-2}$. In fig.5b, the experimental $V_{th}$ data are also compared with the fitting curve given by the equation 5). Best fitting curves have been obtained with β and τ parameters of 0.6 and $1.1*10^4$ sec, respectively. Remarkably, concerning T6 transistors on SiO$_2$ substrates, these values are very similar to those previously found by the same approach and on a comparable time scale[28]. Anyway, in order to analyze and compare the bias stress parameters of different T6 transistors, a different experimental procedure consisting in the direct analysis of current time decay curves upon static polarization (fixed $V_{DS}$ and $V_{GS}$) has been adopted in this work[11,29]. Indeed, combining the equation 5) and the relation 1), another expression accounting directly for the current time decay in the linear regime, where the pinch-off condition is avoided and the traps at organic/dielectric interface are uniformly excited, as a function of β and τ can be easily obtained[11]:

$$I = I_0 \exp[-(\frac{t}{\tau})^{\beta}] \qquad (7).$$

It is significant to stress that in other reports[29] the same expression is used with the addition of a time-independent term (I'), representing the asymptotic current value at infinite time. Anyway, the correct application of the model as described by the equation 5) implies that I=0 at t→∞. Indeed, at t→∞ $V_{th}(t)$= $V_{GS}$ and the current has to be zero if it is assumed to be proportional to ($V_{GS}$ - $V_{th}$) as in relation 1. In general, it is also to be taken into account that the direct observation of the current time decay should be a more reliable criterion for the stress parameter assessment, since any possible ambiguity related to the extraction of the threshold voltage shift from the transfer-curves, in presence of a not constant slope in linear regime, is avoided.

According to this approach, a comparison between current time decay curves referred to transistors with different thickness (d) of the T6 active layer is reported in fig.6a. In any case, the time scale ranged up $10^3$ sec, while the device static operating point has been fixed with $V_{GS}$=-50 V and $V_{DS}$=-5V. Fitting curves according to the relation (7) are also reported. In particular, the fitting procedure has been optimized, with the minimization of the resulting $\chi^2$ as a function of the two stress parameters (β, τ), by using a specialized set of software routines (Minuit)[30].: a) Experimental current time decay curves (scatters) and fitting curves and b) normalized experimental decay curves for T6 transistors with different film thickness.

In fig.6b, the normalized current time decay curves are also presented, clarifying that the film thickness increase affects positively the instability phenomena by the reduction of the current time decay rate. Indeed, while at d=10 nm the bias stress effect reduces the current of a factor higher than 40%, the current decrease is limited to 20% in the case of d=130 nm. This experimental difference is directly reflected in the stress parameters deduced from fitting procedures and summarized in Table 1. Both $\tau$ and $\beta$ parameters resulted considerably thickness dependent and increase with d. The only exception to this trend is given by the value of $\tau$ for d=10 nm where, anyway, charge carriers mobility has been found to be lower than for higher thicknesses, likely pointing at a not complete substrate coverage by the organic layer. It is also noteworthy that the extraction of the $\tau$ value is correlated to the measurement time scales, since different (of a factor between 2 and 3) $\tau$ values have been usually evaluated for the same T6 device by the voltage shift experimental observation and current time decay analysis. Obviously, this discrepancy can be basically related to the different time scales ($10^3$ s vs $2*10^4$ s) adopted in the two experiments, even if, as aforementioned, some uncertainties are associated to the shift voltage determination from the basic transfer-curves. Lower differences (in the range of 20% for the same d) between the two approaches has been experienced for $\beta$ parameter, showing to be likely less sensitive on the measurement time scale. The $\beta$ and $\tau$ values reported in Table 1 are plotted in fig.7 as a function of the film thickness.

Summarizing the results, the organic layer thickness dependence of both stress parameters seems to be at odds with the experimental data (inset of fig.3) about the mobility values, which substantially have been demonstrated to be thickness independent. Anyway, similar experimental evidences have been recently discussed by Chang et al[17] on Pentacene transistors in Bottom-Contact Bottom-Gate configuration. Indicatively, it should be assumed that, although the charge transport is confined in the interfacial region, the energy distribution of the traps involved in free carrier immobilization are affected by film bulk properties with its crystalline structure (i.e. grain boundaries) and not only by the chemical impurities at the gate dielectric/organic interface. In this regard, it is interesting to observe that, by increasing the thickness, the stress and the structural parameters, expressed by X-ray characterization (i.e. rocking curve FWHM), are correlated.

This occurrence seems to indicate that the nature of traps related to bias stress effect differs from that of traps involved in basic charge transport phenomena affecting the mobility. Another possible explanation can be given by invoking the beneficial effect of the increasing film volume in the protection of the first interfacial layers from ambient impurities, whose contamination mainly acts quickly after the deposition. Finally, possible influence of contact resistances can not be completely excluded considering that their contribution to current time evolution under bias stress is neglected in equation 7. In any case, the bias stress dynamics dependence on the film thickness could have a significant technological impact in the organic transistor optimization for practical applications, what motivates future investigation on this specific issue.

## 4 CONCLUSIONS

In this study, bias stress phenomena in T6 transistors have been investigated for films with different thicknesses. The determination and comparison of bias stress parameters, in the framework of a kinetic model suitable to describe the current time decay under static polarization, reveal that the film thickness considerably affect the electrical instability phenomena. In particular, by increasing the organic layer thickness, FET devices seem less sensitive on bias stress effects and stress parameters are correlated with rocking curve structural features. These experimental findings suggest that, differently form the basic charge transport processes related to the carrier mobility, current instability manifestations can be considerably influenced by the bulk properties of the organic semi-conducting film. Finally, our results highlight that the stress parameter extraction procedures must be performed with particular care, mainly focusing on long time scales.

**Acknowledgments**

The authors wish to thank Dr. C. Albonetti, Dr. F. Borgatti and Dr. F. Biscarini for stimulating discussions. The technical support of A. Maggio and S. Marrazzo is gratefully acknowledged.

# FIGURE AND TABLE CAPTURES

Figure 1: Basic transistor configuration and electrode lay-out utilized in this work.

Figure 2: X-rays spectra and rocking curves for devices with different thicknesses.

Figure 3: Comparison between transfer-curves in saturation regime measured in air and vacuum for a T6 Fet. In the inset, mobility extracted in vacuum versus film thickness.

Figure 4: Transfer-curves at different temperatures [from 293K to 70K] for a T6 transistor. In the inset, Arrhenius plot of mobility (µ vs 1/T).

Figure 5: Transfer-curve shift at the $V_{GS}$=-50V bias stress; b) Corresponding threshold voltage shift (scatters) as a function of stress time and the fitting curve (solid line).

Figure 6: a) Experimental current time decay curves (scatters) and fitting curves and b) normalized experimental decay curves for T6 transistors with different film thickness.

Figure 7: Stress parameter β) and (in the inset) τ for T6 transistors a function of the film thickness.

Table 1 Stress parameters for transistors with different T6 thicknesses

**TABLES**

| Film Thickness (nm) | (10 nm) | (30 nm) | (60 nm) | (90 nm) | (133 nm) |
|---|---|---|---|---|---|
| $\beta$ | 0.43 ± 0.01 | 0.50 ± 0.03 | 0.54 ± 0.03 | 0.55 ± 0.03 | 0.67 ± 0.09 |
| $\tau$ | 3727 ± 327 | 3171 ± 261 | 3375 ± 303 | 5933 ± 981 | 8356 ± 2805 |
| $\chi^2$ | 2.19 | 1.33 | 1.71 | 0.70 | 0.20 |

Table 1